# Known-plaintext attack and ciphertext-only attack for encrypted single-pixel imaging

Shuming Jiao,[1] Yang Gao,[1] Ting Lei,[1] Zhenwei Xie,[1] and, Xiaocong Yuan,[1,*]

[1] Nanophotonics Research Center, Shenzhen University, Shenzhen, Guangdong, 518060, China

* xcyuan@szu.edu.cn

Abstract:

In many previous works, a single-pixel imaging (SPI) system is constructed as an optical image encryption system. Unauthorized users are not able to reconstruct the plaintext image from the ciphertext intensity sequence without knowing the illumination pattern key. However, little cryptanalysis about encrypted SPI has been investigated in the past. In this work, we propose a known-plaintext attack scheme and a ciphertext-only attack scheme to an encrypted SPI system for the first time. The known-plaintext attack is implemented by interchanging the roles of illumination patterns and object images in the SPI model. The ciphertext-only attack is implemented based on the statistical features of single-pixel intensity values. The two schemes can crack encrypted SPI systems and successfully recover the key containing correct illumination patterns.

1. Introduction

Optical encryption, authentication and watermarking systems [1-3] can be constructed for information security applications, with advantages such as multi-dimensional parallel processing capabilities, fast processing speed and direct processing of physical objects without digitalization. In previous works, an image encryption system can be physically implemented with various types of optical imaging systems, including but not limited to double random phase encoding (DRPE) [4], holography [5,6], integral imaging [7-9], ptychography [10,11], and single-pixel imaging [12-21]. In these systems, the plaintext image is converted to a light field, which is optically transformed into a ciphertext light field with certain physical encryption keys (e.g. random phase masks or random illumination patterns).

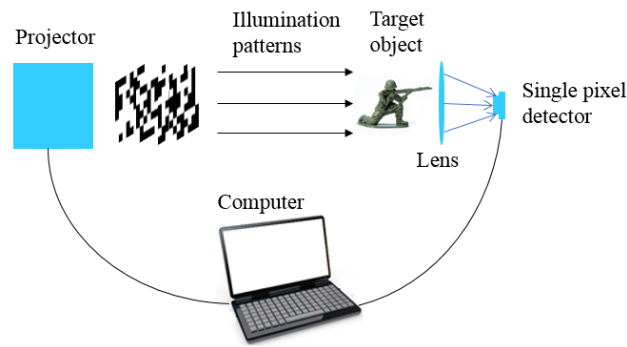

Fig. 1. Optical setup for a single-pixel imaging system.

Single-pixel imaging (SPI) [22,23] is an optical imaging technique that captures an object image with a single-pixel detector instead of a pixelated sensor array. After the target object is sequentially illuminated with many varying patterns and a single-pixel intensity sequence is recorded, the object image can be computationally reconstructed. Compared with other optical encryption architectures [4-11], the sensor is a simple bucket light detector and a real intensity value instead of a complex light field is recorded each time in an encrypted SPI system, which is easier to implement experimentally. A typical optical setup for a SPI system is shown in Fig. 1.

For any type of encryption system, the security strength is always a crucial concern. Attacking methods can be developed to uncover the security flaws of an existing encryption system. In the meanwhile, the security strength of an encryption system can be further enhanced against these attacking methods. Like a shield and spear relationship, encryption methods and attacking methods (or cryptanalysis) are upgraded against each other iteratively to finally produce a more secure system.

The common attacking methods for an image encryption system include chosen-plaintext attack (CPA), known-plaintext attack (KPA) and ciphertext-only attack (COA). In CPA, it is assumed that the attacker can access the encryption system and control the input plaintext content. The security keys are recovered based on selected pairs of plaintexts and ciphertexts. The CPA is relatively easy to implement but the assumption that the attacker can freely choose the plaintext may be invalid in many practical situations. In KPA, a number of plaintext-ciphertext pairs are available, which are given randomly instead of being selected by attackers. KPA can threaten an encryption system under more general conditions compared with CPA. In COA, the attacker can only access a certain number of ciphertexts and does not know any plaintext information. COA requires the least amount of information to crack an encryption system and reveals the most severe security flaw of an encryption system. At the same time, COA is usually most difficult to realize for an attacker.

For other types of optical encryption systems such as DRPE [4], various implementations of encryption systems [24-28] and various types of cryptanalysis [29-35] including CPA, KPA and COA have been extensively investigated. However, for encrypted SPI, many works have been conducted on the design of encryption systems [12-21] since the earliest attempt [12] but little cryptanalysis has been investigated previously. As far as the author knows, only one such work [36] can be found, in which some CPA methods are proposed. No KPA or COA schemes have even been proposed for encrypted SPI systems in the past. In this work, a KPA scheme and a COA scheme for encrypted SPI are investigated for the first time.

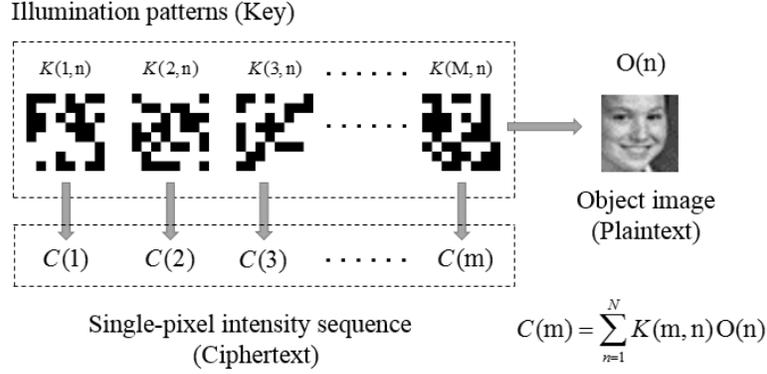

Fig. 2. General framework of an encrypted SPI system.

2. Encrypted single-pixel imaging (SPI) system

In SPI, there are usually three major components, object image $O(n)$ (n=1, 2, ..., N), illumination patterns $K(m,n)$ (m=1, 2, ..., M; n=1, 2, ..., N) and recorded single-pixel intensity sequence $C(m)$ (m=1, 2, …, M). In an encrypted SPI system, the object image $O(n)$ can be employed as the plaintext image, the illumination patterns $K(m,n)$ are employed as the encryption (and decryption) key and the recorded single-pixel intensity sequence $C(m)$ is employed as the ciphertext [12]. The general framework of an encrypted SPI system is shown in Fig. 2.

It is assumed that the total number of pixels in the object image is N and the plaintext image can be represented by a column vector $O(n)$ with length N. The total number of pixels in each illumination pattern is identical to the resolution of the object image. It is assumed that the object image $O(n)$ is sequentially illuminated by a total number of M different illumination patterns and all these patterns jointly constitute an illumination pattern matrix $K(m,n)$ with M rows and N columns. The $m_{th}$ single-pixel intensity value recorded by the detector is the inner product between the $m_{th}$ row in K and the object image $O(n)$ mathematically. The single-pixel intensity sequence can be represented as a column vector $C(m)$ (m=1, 2, ..., M) with length M. The mathematical model of the entire SPI imaging process can be illustrated by Eq. (1). The number of illuminations M can be smaller, equal and larger than the number of pixels N in the object image. The ratio M/N is referred to as the sampling ratio in SPI.

$$\begin{bmatrix} C(1) \\ C(2) \\ \vdots \\ C(M) \end{bmatrix} = \begin{bmatrix} K(1,1) & \cdots & K(1,N) \\ \vdots & \ddots & \vdots \\ K(M,1) & \cdots & K(M,N) \end{bmatrix} \begin{bmatrix} O(1) \\ O(2) \\ \vdots \\ O(N) \end{bmatrix} \quad (1)$$

In SPI, after a total number of M illuminations and recordings, the object image can be computationally reconstructed from $K(m,n)$ ($1 <= m <= M, 1 <= n <= N$) and $C(m)$ ($1 <= m <= M$) with various algorithms [37]. The image reconstruction in SPI is essentially solving a system of linear equations and finding the optimal solutions. As an encryption system, the plaintext image $O(n)$ is recovered when both the ciphertext $C(m)$ and the key $K(m,n)$ are available. For unauthorized users without knowing the key, the plaintext image cannot be disclosed from the ciphertext and its information security is protected.

In practical SPI experiments, the projection devices for pattern illuminations, such as digital micromirror device (DMD), are usually binary [38, 39]. Each pixel in the illumination patterns can be assumed as a random binary value 0 or 1. The total number of possible combinations for M illumination patterns in the key space is $2^{MN}$. A brute-force attack is hard to realize by attackers since it takes a very huge amount of computational resources to attempt all possible combinations to reconstruct the true plaintext image when the key is not known. If any information about the M illumination patterns in the key is not open to public, the system is referred as Type I encrypted SPI system in this paper. The data size of all the M illumination patterns (each one has N pixels) can be considerably large and the transmission & storage cost of encryption (and decryption) keys can be rather high. As an alternative, the permutation of illumination patterns, instead of the original illumination patterns, can be employed as the key, referred to as Type II encrypted SPI system in this paper. In a Type II system, the pixel values of original M illumination patterns before permutation are open to public. However, the order is random and remains secret to authorized users only. For example, the original 5th illumination pattern may be arranged as the first one in the key and the original 7th pattern may be arranged as second one. The total number of possible combinations in the key is $M!$ ("!" refers to factorial) for a total number of M illumination patterns. Even though $M!$ is significantly smaller than $2^{MN}$, the key space in a Type II system still grows very fast as the value of M increases. For example, if there are only M=16 illumination patterns totally, they can already be arranged in $16! = 20922789888000$ possible ways. The number of possible combinations can reach $2.6 \times 10^{35}$ when M=32. A brute-force attack is hard to realize by attackers for a Type II system as well. In Fig. 3, Type I and Type II encrypted SPI systems are compared. In this work, some investigations of KPA and COA are conducted for encrypted SPI systems including both Type I and Type II.

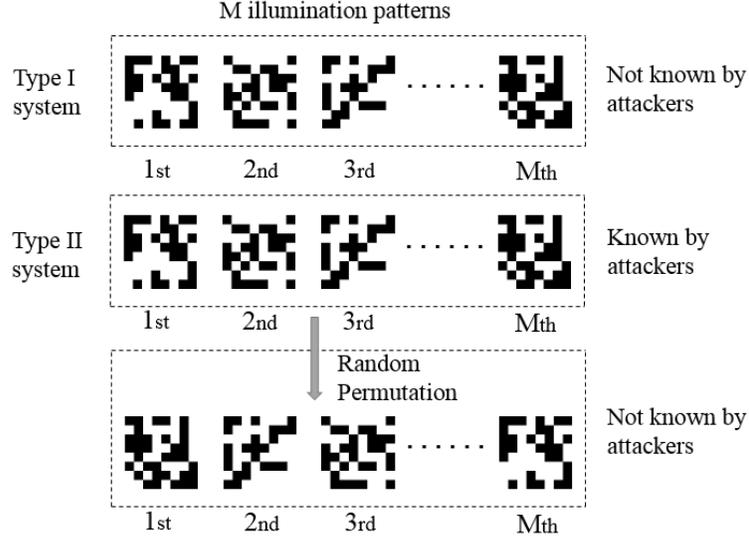

Fig. 3. Type I and Type II encrypted SPI systems.

3. Known-plaintext attack (KPA) to encrypted single-pixel imaging (SPI)

As stated above, the plaintext image can only be recovered from the ciphertext (i.e. the single-pixel intensity sequence) when the key (containing all the illumination patterns) is known. However, if the same key is repetitively employed to encrypt different object images, the attacker may collect all these plaintext images and corresponding ciphertext intensity sequences. From these plaintext-ciphertext pairs, the attacker can possibly figure out most of the pixel values in the key and crack the encryption system. A known plaintext attack (KPA) scheme for encrypted SPI is proposed in this paper for the first time. Our proposed KPA is similar to the image reconstruction process in conventional SPI but the roles of illumination patterns and object images are interchanged. In conventional SPI, the object image is sequentially illuminated by varying illumination patterns. In our proposed KPA model, one illumination pattern is considered to be "sequentially illuminated" by varying plaintext images.

For example, it is assumed that the attacker collects a total number of Q pairs of plaintexts $O_q(n)$ (n=1, 2, ..., N; q=1, 2, ..., Q) and ciphertexts $C_q(m)$ (m=1, 2, ..., M; q=1, 2, ..., Q). For the $m_{th}$ illumination pattern $K_m(n)$ (m=1, 2, ..., M; n=1, 2, ..., N), the following relationship holds, given by Eq. (2).

$$\begin{bmatrix} O_1(1) & O_1(2) & \cdots & O_1(N) \\ O_2(1) & O_2(2) & \cdots & O_2(N) \\ \vdots & \vdots & \ddots & \vdots \\ O_Q(1) & O_Q(2) & \cdots & O_Q(N) \end{bmatrix} \begin{bmatrix} K_m(1) \\ K_m(2) \\ \vdots \\ K_m(N) \end{bmatrix} = \begin{bmatrix} C_1(m) \\ C_2(m) \\ \vdots \\ C_Q(m) \end{bmatrix} \qquad (2)$$

In Eq. (2), each plaintext image $O_q$ [each row in $O_q(n)$ (n=1, 2, …, M; q=1, 2, …, Q) matrix] can be considered as the "$q_{th}$ illumination pattern", the $m_{th}$ illumination pattern $K_m$ can be considered as the "object image" and all the $m_{th}$ elements in the Q ciphertexts can be considered as "intensity sequence" in the KPA model. The $m_{th}$ illumination pattern $K_m$ can be reconstructed from all the plaintext images and all the $m_{th}$ elements in ciphertext single-pixel intensity sequences $C_q(m)$ (q=1, 2, …, Q) with conventional reconstruction algorithms [37] in SPI, shown in Fig. 4 (in comparison with Fig. 2). Each individual illumination pattern in the key is recovered independently from m=1 to m=M. The only difference is that in conventional SPI the object images are usually locally smooth in terms of neighboring pixel intensities and the illumination patterns are orthogonal or random. Total variation (TV) regularization [37,40] can be employed to achieve a high-quality reconstruction with a minimum number of illuminations. However, in KPA, each "object image" is a random illumination pattern and the "illumination pattern" becomes locally smooth when the roles of images and illumination patterns are interchanged. It becomes more difficult to achieve good reconstruction results under a low sampling-rate (a small number of plaintext-ciphertext pairs).

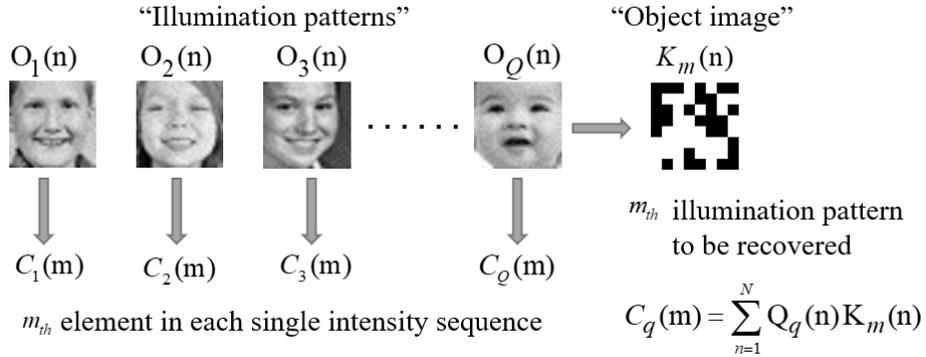

Fig. 4. Proposed KPA model for encrypted SPI.

In this work, the conjugate gradient descent (CGD) algorithm [37,41] is adopted for the recovery of illumination patterns from plaintext-ciphertext pairs. More details about the working principles of this algorithm can be found in the work [37]. In brief, the CGD algorithm is an iterative method to search for an optimal $K_m$ that can best fit Eq. (2). In each iteration, the algorithm calculates the gradient to locate the direction of steepest descent, then performs a line search to locate the optimum step size. The solution moves downhill towards the minimum fitting error efficiently in conjugate directions rather than local gradients. After the optimal $K_m$ is obtained, it will be normalized and binarized.

Each single one in the total of M illumination patterns is recovered individually with the same approach stated above. Finally, the entire encryption key matrix can be recovered and a Type I encryption system is cracked. For a Type II system, we compare each recovered pattern with all the publicly known original patterns and map it to the most similar one that has not been matched previously by other recovered patterns. After the matching steps, the rearranged order of original illumination patterns becomes known and the key can be recovered.

4. Ciphertext only attack (COA) to encrypted single pixel imaging (SPI)

In KPA, multiple pairs of plaintext images and ciphertext intensity sequences are known and the illumination patterns (encryption / decryption key) are recovered. In COA, only a certain number of single-pixel intensity sequences are available and it is a very challenging task to directly recover the illumination patterns. In this work, only the COA for a Type II encrypted SPI system under certain conditions is attempted. It is assumed that SPI is performed repetitively to the same category of images (e.g. different handwritten numbers). The attacker cannot access any of the original object images (plaintexts). However, the attacker may collect a large number of exemplar images similar to the actual object images.

The basic idea is that the single-pixel intensity values recorded with the same illumination pattern for the same category of object images follow a certain statistical distribution. The attacker may be able to figure out the illumination patterns by comparing the histograms of single-pixel intensity values generated from the actual object images with the ones virtually generated from the exemplar object images he has collected.

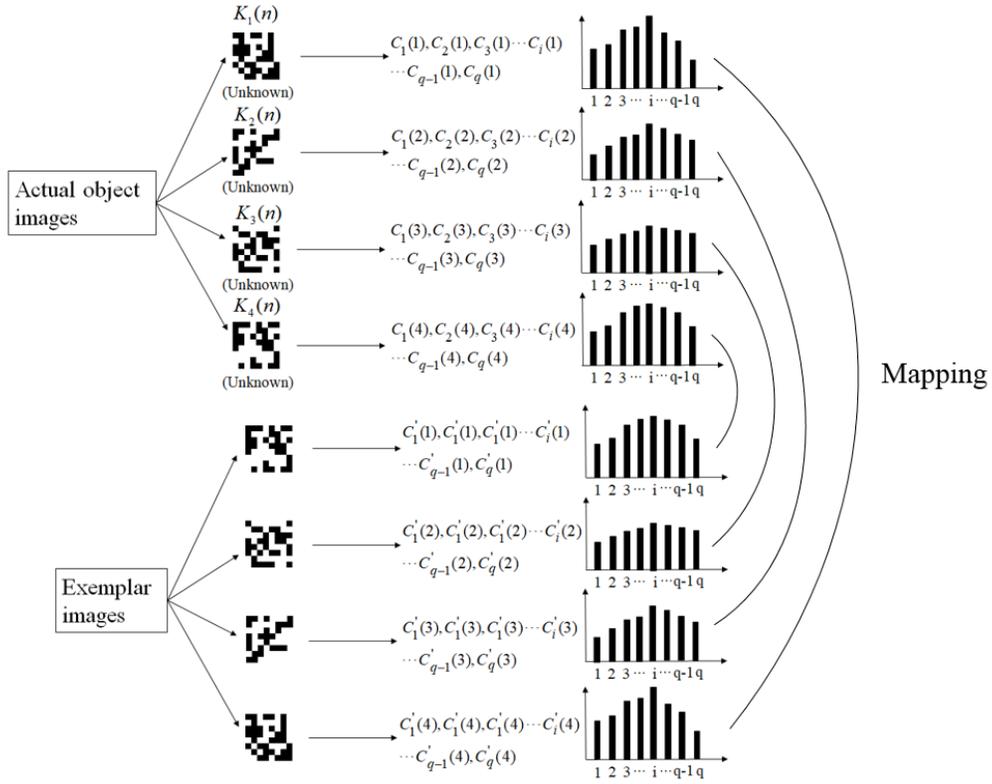

Fig. 5. Proposed COA model for encrypted SPI.

For a Type II system, the goal of attacking is to find the order how the original sequence of illumination patterns is permutated. From Q object images, Q single pixel intensity sequences $C_q(m)$ (q=1, 2, …, Q) can be recorded with the $m_{th}$ rearranged illumination pattern in the key. The attacker also collects Q exemplar images, then Q single pixel intensity sequences $C'_q(m)$ (q=1, 2, …, Q) can be virtually generated with the $m_{th}$ original illumination pattern. The attacker can generate the intensity sequences $C'_q(1)$, $C'_q(2)$, …, $C'_q(M)$ (q=1,2,..,Q) for all the M original illumination patterns. For the $m_{th}$ illumination on the actual object images, the attacker compares the statistical distribution of $C_q(m)$ with $C'_q(1)$, $C'_q(2)$, … $C'_q(m)$ and (q=1, 2, …, Q) and find the most matched one. For example, if $C_q(1)$ is most similar to $C'_q(5)$, then the attacker can conclude the $5_{th}$ original illumination pattern is arranged as the 1st one in the encryption key. After a mapping for each $C_q(m)$ to $C'_q(m)$ (m=1,2, …, M), the attacker can obtain the correct permutation of the illumination patterns and recover the key.

Mathematically, the similarity in the statistical distributions between [ $C_1(m)$, $C_2(m)$, ..., $C_Q(m)$ ] (m=1,2,...,M) and [ $C_1'(m')$, $C_2'(m')$, ..., $C_Q'(m')$ ] ($m'$=1, 2, ..., M) can be evaluated in the following way. The intensity values of $C_q(m)$ and $C_q'(m)$ (q=1, 2, ..., Q) are statistically distributed within a certain range (e.g. [0 20]). The range can be divided into uniform intervals with certain bin size. For example, if the bin size is 5, there will be four intervals [0 5], [5 10], [10 15] and [15 20]. Then the number of intensity values falling into each interval is counted and a vector containing these numbers (e.g. [7 13 16 8]) is obtained. The difference between two number-count vectors (i.e. histogram) can be measured by their Euclidean distance. $C_q(m)$ will be mapped to $C_q'(m')$ if the Euclidean distance between the number-count vectors of $C_q(m)$ and $C_q'(m')$ is smallest, indicating that their statistical distributions are most similar.

## 5. Results and Discussions

### 5.1 Known-plaintext attack (KPA)

In the simulation of KPA, the number of pixels in the object image and in each illumination pattern is assumed to be N= $32 \times 32$. Three different numbers of binary random illumination patterns, corresponding to three different sampling ratios $M/N$ =0.4, $M/N$ =0.7 and $M/N$ =1, are tested in an Type I encrypted SPI system. The object image is reconstructed with total variation (TV) regularization algorithm [37,41] from the illumination patterns and single-pixel intensity sequences. It is assumed that a varying number of plaintext images and corresponding ciphertext intensity sequences are available to the attacker. The plaintext images are all human face images taken from UTKFace dataset [42], with some examples shown in Fig. 6(a). The illumination patterns, as the encryption and decryption key, are not known by the attacker originally. From these available plaintext-ciphertext pairs, the attacker can crack the encryption system and recover the illumination patterns with the KPA method stated in Section 3.

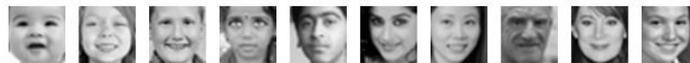

(a)

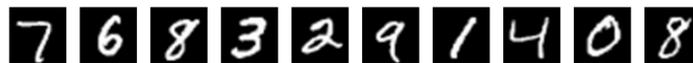

(b)

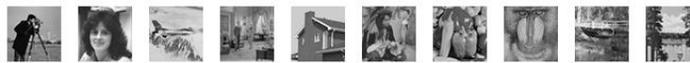

(c)

Fig. 6. Object images in simulation: (a)Examples of plaintext images in UTKFace dataset [39]; (b) Examples of plaintext images in MISNT database [40]; (c)Ten testing images for KPA.

The accuracy of attacking results is evaluated in two ways. First, all the binary pixels in the illumination patterns of the correct key and in the cracked illumination patterns after attacking are compared and the cracking correct rate (percentage of correct pixels) is calculated. Second, ten testing images shown in Fig. 6(c), which are completely different from the plaintext images used for attacking, are encrypted by the patterns in the correct key and then decrypted with both the correct key and the cracked key using our proposed KPA scheme. The average peak signal-to-noise ratio (PSNR) values are calculated and compared for these two cases. The attacking results for a Type I encrypted SPI system are shown in Table 1 and some examples of final recovered images are shown in Fig.7.

Table 1. Results of our proposed KPA scheme for Type I encrypted SPI system.

| N | M ($M/N$) | Q ($Q/N$) | Cracking Correct Rate |
|---|---|---|---|
| 32×32 | 410 (0.4) | 1024(1) | 0.9668 |
| | | 2048(2) | 0.9935 |
| | | 3072(3) | 0.9978 |
| | | 4096(4) | 0.9991 |
| 32×32 | 717 (0.7) | 1024(1) | 0.9655 |
| | | 2048(2) | 0.9957 |
| | | 3072(3) | 0.9974 |
| | | 4096(4) | 0.9983 |
| 32×32 | 1024 (1) | 1024(1) | 0.9683 |
| | | 2048(2) | 0.9944 |
| | | 3072(3) | 0.9976 |
| | | 4096(4) | 0.9982 |

It can be observed from the results in Table 1 and Fig. 7 that the pixels in the illumination patterns can be recovered with very high accuracy (e.g. over 99.8%) and the test images can be reconstructed with acceptable visual quality (e.g. a PSNR of over 16dB) using the recovered key, if the attacker has a sufficient number of plaintext-ciphertext pairs. In Fig. 7(a), the decrypted (or reconstructed) images with the correct key have good fidelity. The decrypted images with random wrong keys appear like noise and no information about the actual object image can be visually perceived, shown in Fig. 7(b). This indicates that an encrypted SPI system exhibits a substantial level of security when the key is not known. However, after our proposed KPA is performed, the decrypted images with the recovered key are close to the original plaintext images, if the number of plaintext-ciphertext pairs available is adequate, shown in Fig. 7(c), Fig. 7(d) and Fig. 7(e).

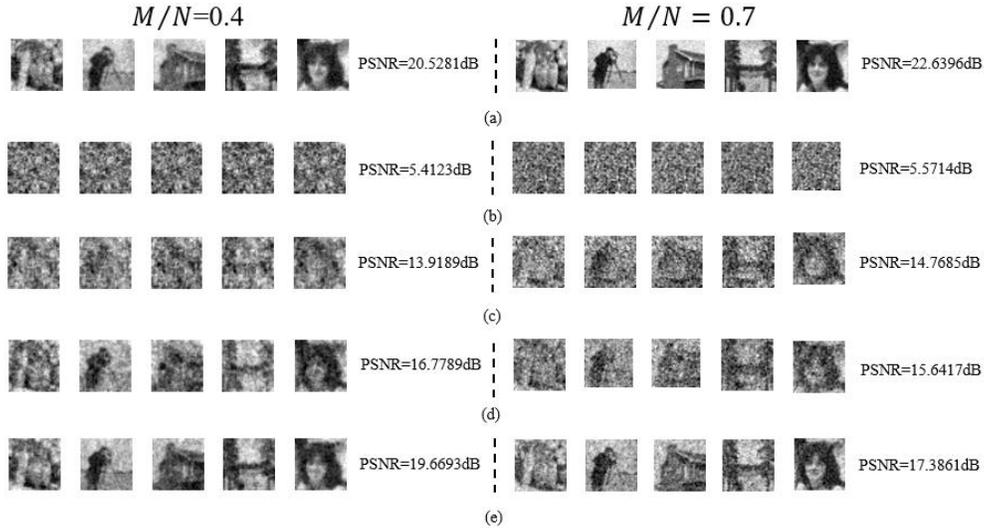

Fig. 7. KPA results when $N = 32 \times 32$ and the sampling rate $M/N$ is 0.4 (left) or 0.7 (right) for a Type I encrypted SPI system: (a) Decrypted results with correct key; (b) Decrypted results with random wrong key; (c) Decrypted results with recovered key when q=2048; (d) Decrypted results with recovered key when q=3072; (e) Decrypted results with recovered key when q=4096.

The results reveal that our proposed KPA scheme is an effective approach to crack a Type I encrypted SPI system. It can be observed that when most of the pixels in the illumination patterns are successfully recovered (e.g. the correct rate is around 97%), the reconstructed object image using recovered patterns will still be rather poor (e.g. 12dB or 13dB) due to the small percentage of error bits. The encryption system can be truly cracked only when the illumination patterns are recovered with very high accuracy (e.g. over 99.6%) for complicated grayscale plaintext images [like Fig. 6(c)]. Since each illumination pattern in the key is recovered independently one by one in our scheme, the number of plaintext-ciphertext pairs required for successful attacking will not evidently increase when the sampling ratio $M/N$ increases. However, the number of plaintext-ciphertext pairs required for successful attacking is relevant to the number of pixels N (or image size). The relationship between cracking correct rate and image size is demonstrated in Fig. 8. The results indicate that more plaintext-ciphertext pairs will be required for cracking one illumination pattern containing more pixels.

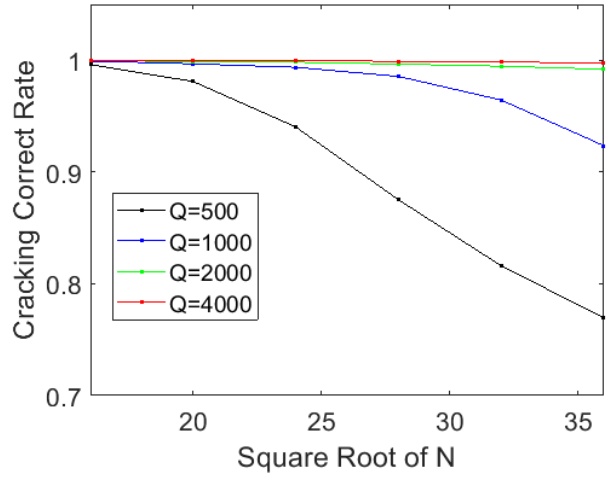

Fig. 8 Performance of our proposed KPA scheme for a Type I encrypted SPI system when the image size varies (Sampling rate M/N=0.7)

Our proposed KPA scheme is also implemented for a Type II encrypted SPI system. The parameters in the simulation are basically the same as above except that the number of plaintext-ciphertext pairs attempted is much smaller. The results are demonstrated in Table 2 and Fig. 9.

Table 2. Results of our proposed KPA scheme for a Type II encrypted SPI system.

| N | M (M/N) | Q (Q/N) | Cracking Correct Rate |
|---|---|---|---|
| 32×32 | 410 (0.4) | 31 (0.03) | 0.9434 |
|  |  | 51 (0.05) | 0.9929 |
|  |  | 72 (0.07) | 0.9977 |
|  |  | 92 (0.09) | 1 |
| 32×32 | 717 (0.7) | 31 (0.03) | 0.9219 |
|  |  | 51 (0.05) | 0.9972 |
|  |  | 72 (0.07) | 0.9986 |
|  |  | 92 (0.09) | 1 |
| 32×32 | 1024 (1) | 31 (0.03) | 0.8975 |
|  |  | 51 (0.05) | 0.9936 |
|  |  | 72 (0.07) | 0.9980 |
|  |  | 92 (0.09) | 1 |

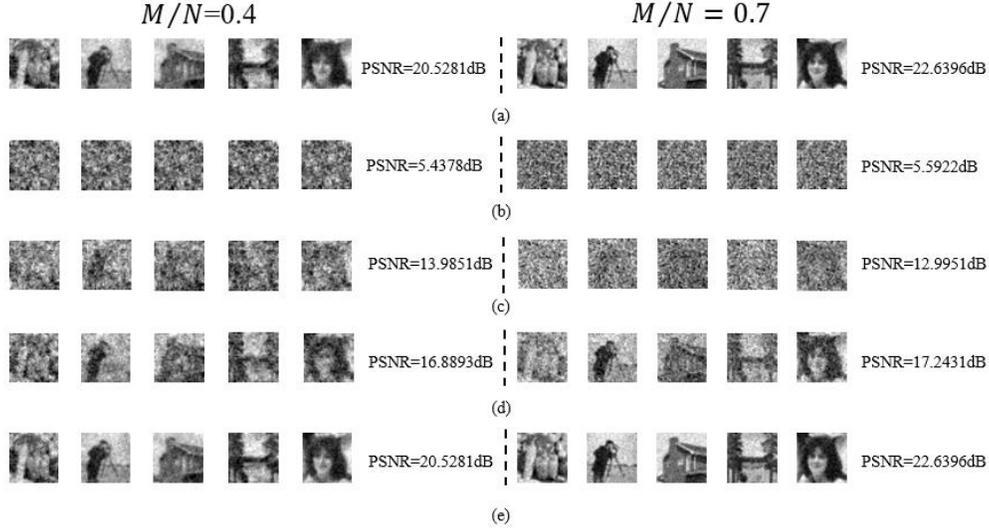

Fig. 9. KPA results when N=$32\times32$ and the sampling rate ($M/N$) is 0.4 (left) or 0.7 (right) for a Type II encrypted SPI system: (a) Decrypted results with correct key; (b) Decrypted results with wrong key; (c) Decrypted results with recovered key when q=31; (d)Decrypted results with recovered key when q=72; (e) Decrypted results with recovered key when q=92.

From the results above, it can be observed that our proposed scheme can be utilized to crack a Type II system as well. The number of plaintext-ciphertext pairs required for a successful attack to a Type II system is much smaller than a Type I system. As stated in Section 2, the number of possible combinations in the key space is $2^{MN}$ for a Type I system and M! for a Type II system. The attacker needs to recover every pixel in the illumination pattern for a Type I system but only needs to recover the pattern permutation for a Type II system. The security strength of a Type II system is significantly lower and therefore relatively easy to be cracked. When the entire order of permutation is fully recovered (e.g. when Q=92 in Table 2), all the original illumination patterns can be rearranged correctly and 100% pixels in the recovered key will have correct values (correct rate=1). In comparison, for a Type I system, the majority of pixel values can be recovered with our proposed KPA scheme but it is hard to achieve 100% correct rate.

*5.2 Ciphertext-only attack (COA)*

In the simulation of COA, the number of pixels in the object image and in each illumination pattern is assumed to be N= $8\times8$, $12\times12$ and $16\times16$. The number of binary random illumination patterns M is assumed to be 64. The plaintext images are resized digital number images taken from the MISNT database [43], shown in Fig. 6(b). It is assumed that there are totally Q plaintext images and Q corresponding single-pixel sequences. The Q single-pixel sequences are the only available data to an attacker.

A Type II encrypted SPI system is considered and the 64 original illumination patterns are randomly permutated by rearranging the order. In addition, it is assumed that the attacker can collect Q exemplar images similar to the actual plaintext images. It shall be noted that the exemplar images also belong to the MINST database but they are different from the Q plaintext images. Both the plaintext images and exemplar images are randomly chosen from the MINST database. Since our proposed COA scheme is based on statistical features, the number Q has to be sufficiently large, governed by the law of large numbers. In the simulation, Q is set to be 6000, 10000 and 14000. In the comparison of single-pixel intensity distributions, the intensity value range is [0 15] and the bin size is 0.5. The COA results are shown in Table 3 and Fig. 10.

Table 3. Results of our proposed COA scheme for a Type II encrypted SPI system.

| N | M | Q | Cracking Correct Rate |
|---|---|---|---|
| 8×8 | 64 | 6000 | 0.8854 |
|  |  | 10000 | 0.9427 |
|  |  | 14000 | 1 |
| 12×12 | 64 | 6000 | 0.7813 |
|  |  | 10000 | 0.9063 |
|  |  | 14000 | 1 |
| 16×16 | 64 | 6000 | 0.5469 |
|  |  | 10000 | 0.6875 |
|  |  | 14000 | 0.8438 |
| 32×32 | 64 | 6000 | 0.03125 |
|  |  | 10000 | 0.046875 |
|  |  | 14000 | 0.0625 |

From Table 3, it can be observed that our proposed COA scheme can fully crack a Type II encrypted SPI system when the image size is sufficiently small, and the number of plaintexts Q is sufficiently large. The order of rearranged illumination patterns in the key can be possibly fully recovered (correct rate=1). As the image size grows, the performance of our proposed COA scheme will become degraded but it can still partially recover the key. In Fig. 10(a), several 8×8 English letter images are used as the testing plaintext images. After these images are encrypted with the original key, they can be successfully decrypted with our recovered key, shown in Fig. 10.

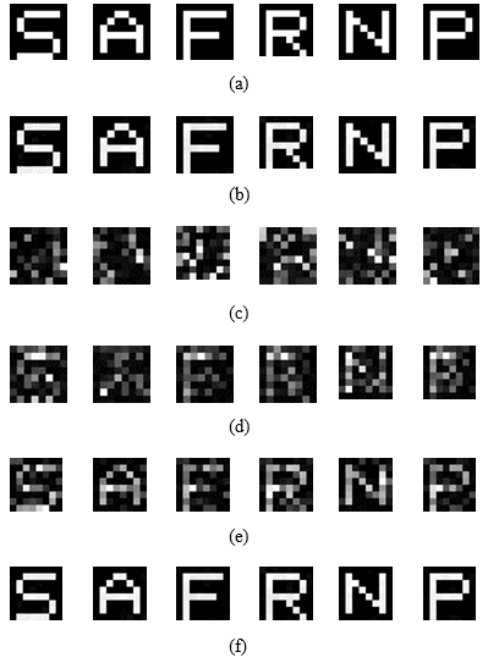

Fig. 10 COA results when the N=8×8 and M=64 for a Type II encrypted SPI system: (a) Original plaintext images; (b) Decrypted results with correct key; (c) Decrypted results with wrong key; (d) Decrypted results with recovered key when q=6000; (e) Decrypted results with recovered key when q=10000; (f) Decrypted results with recovered key when q=14000.

In Fig. 11, the histograms of recorded single-pixel intensity values from 14000 actual plaintext images and 14000 exemplar images collected by the attacker for three different illumination patterns are presented. The distributions of single-pixel intensities generated with the same illumination pattern for two different object image sets (belonging to the same image category) are quite similar statistically. But the intensity value distributions between different patterns are obviously different. This important property is utilized for matching the unknown pattern with a list of original patterns to recover the order of permutation.

Our proposed COA scheme can perform well under some circumstances but may fail under other circumstances, due to the following reasons. First, all the plaintext images shall be similar and belong to the same category. Ideally, the intensity value at each pixel position follows a unique statistical distribution. In this way, the single-pixel intensity values will exhibit different statistical features when these images are illuminated by different random patterns. If the plaintext images are random and have almost no common features in the pixel intensity distribution, it is hard to observe distinguishable statistical features in the recorded single-pixel intensity values for varying illumination patterns. Second, the attacker needs to collect an adequate number of exemplar images similar to the actual images. Ideally, the intensity value in each corresponding pixel in the actual images and exemplar images follows the same statistical distribution. In our simulation, when the exemplar images collected by the attacker are replaced with human face images, instead of number digit images, none of the 64 patterns can be recovered in the correct order (N= $8\times8$, M=64 and Q=14000). Third, as shown in the results above, our proposed COA scheme can work better for plaintext images with a smaller number of pixels. It may give poor attacking results or even completely fail to work when N is large. Since the single-pixel intensity is a weighted sum of all the pixel values in one plaintext image, the features in the intensity distributions for different pixels in the plaintext images are more likely to be lost when many pixel values are combined into a single-pixel intensity value.

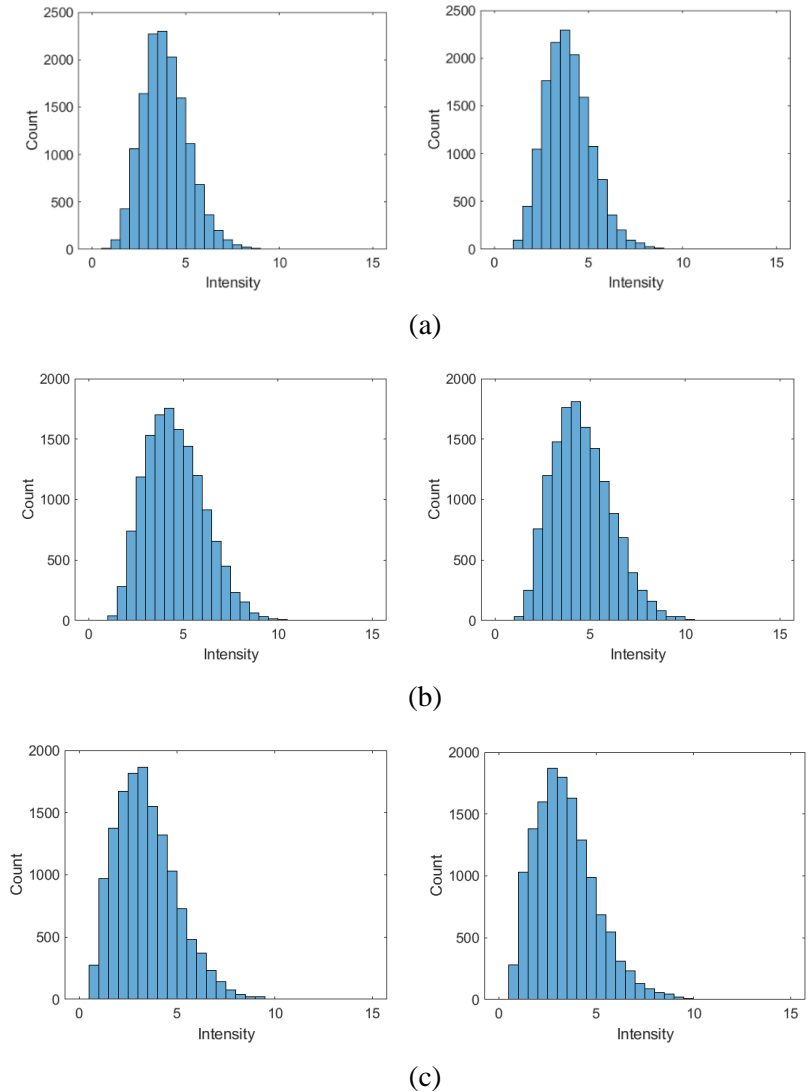

Fig. 11. Examples of the histograms for recorded single-pixel intensity values from 14000 actual plaintext images (left) and 14000 exemplar images collected by the attacker (right) in COA for three different illumination patterns: (a) Pattern 1; (b) Pattern 2; (c) Pattern 3.

6. Conclusion

In many previous works, a single-pixel imaging (SPI) system is constructed as an optical image encryption system. In such a system, the object image is employed as the plaintext, the illumination pattern is employed as the encryption & decryption key and the recorded single-pixel intensity sequence is employed as the ciphertext. Like a shield and spear relationship, encryption methods and attacking methods play an equally critical role in the investigation of a certain type of cryptosystem. Most previous works focus on the design of an encrypted SPI system but little corresponding cryptanalysis has ever been conducted. In this work, we propose a known-plaintext attack (KPA) scheme and a ciphertext-only attack (COA) scheme to an encrypted SPI system for the first time. The KPA is implemented by interchanging the roles of illumination patterns and object images in a SPI model. The secret illumination patterns can be recovered from a set of known plaintext images and ciphertext intensity sequences. The COA is implemented based on the statistical features in the distribution of single-pixel intensity values when the SPI is performed on the same category of images. The attacker can compare the statistical distributions of single-pixel intensity values generated from the actual object images and virtually generated from the collected exemplar images. The unknown illumination patterns can be recovered after being mapped to known original patterns. Simulation results verify the effectiveness of our proposed attacking schemes. The two schemes can crack encrypted SPI systems and successfully recover the key containing correct illumination patterns. Our proposed attacking schemes reveal some security flaws of an encrypted SPI system under certain circumstances. In addition, our proposed attacking algorithms may be meaningful for other applications similar to the cryptoanalysis of an optical encryption system, like scattering imaging.

Funding

National Natural Science Foundation of China (61805145, 11774240); Leading talents of Guangdong province program (00201505); Natural Science Foundation of Guangdong Province (2016A030312010); Science and Technology Innovation Commission of Shenzhen (KQJSCX20170727100838364)